\documentstyle[psfig,aas2pp4]{article}

\slugcomment{to appear in the Astrophysical Journal 2002} 
 
\begin{document}
 
\title{HI 21cm absorption beyond the epoch of re-ionization}
   
\author{C. L. Carilli}
\affil{National Radio Astronomy Observatory, P.O. Box O, Socorro, NM,
87801, USA \\
ccarilli@nrao.edu}
\author{N.Y. Gnedin}
\affil{Center for Astrophysics and Space Astronomy, 
University of Colorado, Boulder, CO 80309}
\author{F. Owen}
\affil{National Radio Astronomy Observatory, P.O. Box O, Socorro, NM,
87801, USA}

\begin{abstract}

We explore the possibility of detecting HI 21cm absorption by the
neutral intergalactic medium (IGM) toward very high redshift radio
sources.  The epoch considered is between the time when the first
ionizing sources form and when the bulk of the neutral IGM becomes
ionized. Due to the extreme Ly$\alpha$ opacities of the neutral IGM,
objects within this 'gray age' can only be observed at wavelengths
longer than about 1$\mu$m.  We use the latest simulations of the
evolution of the IGM in the context of $\Lambda$CDM structure
formation models constrained by observations of the highest redshift
QSOs to predict the optical depth as a function of frequency of the
neutral IGM due to the HI 21cm line. We then simulate radio spectra
assuming observational parameters for future large area radio
telescopes.  These spectra show that HI 21cm absorption studies could
be a powerful probe of the rich structure of the neutral IGM prior to
the epoch of reionization, including $\sim 1\%$ absorption by the mean
neutral IGM, plus deeper, narrow lines ($\ge 5\%$ and a few km
s$^{-1}$). 
Most of the variations in transmissivity are due to the mild density
inhomogeneities with typical values of the cosmic overdensity
$\delta\sim 10$, precisely the structures that at later times give rise to
the Ly$\alpha$ forest.
We also consider sensitivity limits and the evolution of
radio source populations, and conclude that it is reasonable to
hypothesize the existence of an adequate number of high-$z$ radio
sources against which such absorption studies could be performed.

\end{abstract}
 

\section {Introduction}

After recombining at $z \sim 1000$, the intergalactic medium
(IGM) remains neutral until the formation of the
first luminous objects. Standard $\Lambda$CDM structure formation
models suggest that the first objects with sufficient mass
to ionize substantially the IGM 
form at $z \le 15$ (Gnedin \& Ostriker 1997).  
It is currently not known whether these
first luminous objects are active star forming galaxies or
accreting massive black holes, but given the current observations
of both galaxies and luminous AGN at $z > 6$  (Hu et al. 2002; Fan et
al. 2001), it seems likely that a mixture of sources exists.
Reionization of the IGM proceeds slowly at first, with each
object essentially contained within its own Stromgren sphere. 
Eventually these spheres overlap such that ionizing photons
from a given object contribute to multiple spheres, resulting
in a run-away process. 
This overlap phase is known as 'fast' reionization, and in the
discussion below we call this phase the 'epoch of reionization'
(Loeb \& Barkana 2001; Gnedin 2000; 2002).

The recent detection of Ly$\alpha$ absorption by the neutral IGM in
two $z \sim 6$ QSOs, as predicted by Gunn \& Peterson (1965), has
revolutionized our understanding of cosmic reionization by placing
constraints on the epoch of reionization (Becker et al. 2002;
Djorgovski et al. 2002; Pentericci et al. 2002).  Using these
constraints in the context of $\Lambda$CDM structure formation models,
Gnedin (2002) fixes the redshift of fast reionization at $z \sim 6.2$,
although admittedly the possibility of cosmic variance makes this
estimate uncertain (Hu et al. 2002).  

The picture then is one of a 'gray-age' between the epoch of formation
of the first luminous objects and the epoch of fast
reionization. During 
the gray-age the IGM is predominantly neutral, with pockets of ionized
gas around the first luminous objects. The neutral IGM is opaque at
rest wavelengths shorter than Ly$\alpha$ such that study of objects,
and the IGM, during this age will be limited to observations at
wavelengths longer than about 1$\mu$m.  One method for studying
structures during this gray-age is through the HI 21cm line of neutral
hydrogen. Many groups have considered observations of HI 21cm
emission from the neutral IGM during the gray-age, 
or absorption against the microwave background (depending on
the HI excitation temperature; Scott \& Rees 1990; Bagla,
Nath, \& Padmanabhan 1997; Tozzi et al. 2000; 
Shaver et al. 1999; Iliev et  al 2002; Bharadwaj \& Sethi 2002).
However, even in the most optimistic case of 
thermal noise limited observations, 
the sensitivity of large area
future radio telescopes, such as the Square Kilometer Array (SKA),
implies that such studies will be limited to 
large scale structure (M(HI) $> 10^{12}$ M$_\odot$). 
Including systematic errors, such as source confusion
(di Matteo et al. 2001), could degrade substantially this limit.

In this paper we consider studying the neutral IGM
during the gray-age via HI 21cm absorption studies
toward discrete radio sources. The important point is that
while the IGM is opaque to the Ly$\alpha$ line,
the weakness of the magnetic hyperfine transition
makes the IGM translucent to HI 21cm absorption. 
Column density sensitivity for absorption studies is set
only by the surface brightness of the background source, thereby 
allowing absorption studies to probe to orders-of-magnitude lower
masses than can be detected in emission.  

We use models of structure formation and cosmic
reionization to predict the HI 21cm optical depth ($\tau$) of the IGM
during the gray-age.  We then address the questions: 
Are future radio telescopes adequate to detect these signals?  And 
are there high redshift radio sources of sufficient 
luminosity to permit these studies?  
In a future paper we will consider in more detail the
physical information about the rich structures in the pre-reionization
IGM that can be obtained through HI 21cm absorption observations. We
assume $H_0$ = 70 km s$^{-1}$ Mpc$^{-1}$, $\Omega_M = 0.35$, and
$\Omega_\Lambda = 0.65$. 

\section{Models for the HI 21cm optical depth
of the  IGM during the gray-age}

Our analysis relies on the simulations of Gnedin (2000, 2002). 
These simulations include the three
main physical ingredients required to model neutral hydrogen
absorption in the redshifted 21 cm line: inhomogeneous small-scale
structure of the universe, radiative transfer, and accurate treatment
of the level populations in atomic hydrogen.
The dynamics of gas and dark matter is followed in a quasi-Lagrangian
fashion (i.e.\ with high resolution) with the Softened Lagrangian
Hydrodynamics (``SLH'') code (Gnedin 1995, Gnedin \& Bertschinger
1996). The radiative transfer is modeled with the newly developed
Optically Thin Variable Eddington Tensor (OTVET) approximation (Gnedin
\& Abel 2001). Finally, we include all the effects that couple gas
kinetic temperature to the spin temperature of the atomic hydrogen:
Ly$\alpha$ pumping and collisions with electrons and neutral atoms.
The simulations are normalized in such a way as to reproduce both the
observed star formation rate at $z\sim4$ (Steidel et al.\ 1999) and
the observed evolution of the mean transmitted flux in the spectra of
two $z\sim6$ quasars (Becker et al.\ 2001; Djorgovski et al.\ 2001;
Pentericci et al.\ 2002; Fan et al.\ 2002). 
We can therefore expect
that these simulations are at least representative of the physical
processes involved,  with the caveat that some of the details, such as
the exact redshift of fast reionization, remain uncertain due to the
paucity of constraints.

Figure 1 shows the evolution of the volume averaged kinetic and spin
temperatures and the CMB temperature over a range of interesting
redshifts. Tozzi et al. (2000) showed that 
during the gray-age the excitation temperature of the
HI (the 'spin temperature', $T_S$), is likely to be in excess of the
CMB temperature due to the standard
Wouthuysen-Field effect, i.e.\ resonant scattering of the ambient
Ly$\alpha$ photons emitted by the first ionizing sources (at higher
densities
collisions with electrons and neutral atoms also play a role).
As the structure develops, the kinetic temperature $T_K$
increases both due to
shock heating of the gas in high density regions and Ly$\alpha$ heating in
low density regions. However, as the Ly$\alpha$ excitation rate $P_\alpha$
does not
increase very fast with redshift, the spin temperature 
\begin{equation}
T_S = T_{\rm CMB}{P_\alpha+P_{\rm th}\over P_{\rm th}+P_\alpha 
T_{\rm CMB}/T_K}
\end{equation}
(here $P_{\rm th}=7.6\times10^{-13}{\rm s}^{-1}(1+z)$ is the so-called
``thermalization `` rate, Madau, Meiksin, \& Rees 1997, Tozzi et al.\ 2000)
increases at a
slower rate, because the spin temperature becomes independent of the
kinetic temperature when the latter gets very large.

Figure 2 shows physical quantities along a representative line of
sight through the simulation box at three different redshifts
($z = 8, 10, 12$), with redshift indicated by the line color. 
The top panel shows the HI 21cm
transmissivity of the neutral IGM at high velocity resolution (0.5 kHz
$= 1$ km s$^{-1}$ at 150 MHz).  The abscissa
for this panel is velocity. The middle panel shows the kinetic
and spin temperatures of the gas.  
The bottom panel shows the neutral hydrogen density
structure. The abscissa in these two cases is 
the  corresponding comoving physical scale.
Most of the variations in transmissivity are due to the mild density
inhomogeneities with typical values of the cosmic overdensity
$\delta\sim10$, precisely the structures that at later times give rise to
the Ly$\alpha$ forest. 
Because these structures are typically filamentary,
they are at first shock-heated to about 100 K. At $z<10$
resonant Ly$\alpha$ scattering further increases the gas temperature to
several hundred degrees. At the same time the first HII regions start to
appear - one of them manifests itself in the sharply lower neutral hydrogen
density at the right edge of the bottom panel at $z=8$. 

The optical depth, $\tau$, of the neutral hydrogen to 21cm absorption
for our adopted values of the cosmological parameters is
\begin{equation}
  \tau = 0.008 \left(T_{\rm CMB}\over T_S\right)
  \left(1+z\over 10\right)^{1/2}x_{\rm HI}(1+\delta),
  \label{tauofz}
\end{equation}
where $x_{\rm HI}$ is the neutral hydrogen fraction, and $\delta$ is
the cosmic overdensity (Tozzi et al.\ 2000).  Figure 3 shows in two
panels the joint distribution of the spin temperature, $T_S$, and
neutral hydrogen density (in units of the mean hydrogen density),
$x_{\rm HI}(1+\delta)$, of gas along random lines of sight at two
redshifts: $z=12$ and $z=8$. These parameters, plus $T_{\rm CMB}$,
dictate the HI 21m optical depth via equation 2.  The solid lines in
Figure 3 are iso-$\tau$ curves, for $\tau = 0.002$, 0.01, and 0.05.
At $z = 12$ most of the IGM has $\tau \sim 1\%$, and some of the high
density regions reach $\tau \sim 5\%$ and above (a finite size of the
computational box limits our ability to trace the high $\tau$ tail of
the distribution).  These higher $\tau$ points typically have
relatively narrow velocity widths, as low as a few km s$^{-1}$ 
(Figure 2), implying HI column densities of order 10$^{19}$ to 
10$^{20}$ cm$^{-2}$.
By $z \sim 8$, Ly$\alpha$ heating of the low density gas increases
the mean spin temperature to above 100 K and the mean IGM optical
depth has dropped to $\tau \sim 0.1\%$. However, narrow, higher $\tau$
absorption lines that form in the still neutral
filaments are still easily identifiable.

The deeper, narrow lines occur over the entire redshift range from 6
to 15, but are more prevalent at higher redshift.  The redshift
density of lines with $\tau \ge 2\%$ at $z = 10$ is about 50 lines
per unit redshift, while at $z = 8$ this decreases to about 4 lines
per unit redshift.  We note that at $z \sim 4$, 
well below the epoch of
reionization, the redshift density of high column density
absorption line systems (ie. the damped Ly $\alpha$ systems with N(HI)
$\ge 10^{20}$ cm$^{-2}$) is about 0.3 per unit redshift
(Storrie-Lombardi \& Wolfe 2000).  These latter systems are thought to
arise in gas associated with (proto-) galactic disks and/or halos
(Wolfe \& Prochaska 2000).

In general, the spectra shown in Figure 2 show that the structure of
the neutral IGM during the gray-age is rich in both temperature,
density, and velocity structure. Studying this structure via HI 21cm
absorption lines will offer important clues to the evolution of the
IGM at the very onset of galaxy formation.

\section{Simulated spectra of high redshift radio sources}

\subsection{Telescope parameters}

Plans for future large radio telescopes involve building an aperture
synthesis array with roughly a square kilometer of collecting area,
namely a 'square kilometer array'
(SKA\footnote{http://www.skatelescope.org}).  The detailed design and
specifications for such an array are currently being considered.  In
the analysis below we assume an effective area of $5\times10^5$ m$^2$
at 200 MHz, two orthogonal polarizations, and a system temperature of
250 K (100 K from the receiver and 150 K from diffuse Galactic
emission\footnote{The contribution to system temperature
from diffuse Galactic nonthermal emission 
behaves as frequency$^{-2.75}$  in this frequency range.}).  We also make
the simplifying assumption that the ratio of 
effective area to system temperature remains roughly constant down to
100 MHz, as could arise in the case of a low frequency array composed
of dipole antennas, and hence that the sensitivity is constant across
the frequency range of interest (100 MHz to 200 MHz).  We adopt a
long, but not unreasonable, integration time of 10 days (240
hours). These parameters lead to an expected rms noise level of
34$\mu$Jy in a 1 kHz spectral channel.  
In section 3.3  we assume thermal (ie. Gaussian) noise limited spectra.
A Gaussian noise generator based on algorithms in the
AIPS software and using the standard  FORTRAN random number generator 
is used, normalized to an rms level of 34$\mu$Jy.

We assume that the
correlator will provide at least 10$^4$ spectral channels over a 10
MHz band, implying a channel width of 
1 kHz = 2 km s$^{-1}$.
We also assume that the spectral bandpass determination will be at
least as  good as current telescopes, and presumably considerably
better. Current telescopes such as the VLA and the WSRT can attain
spectral dynamic ranges of up to 10$^4$ over 10's of MHz of radio
spectrum with proper attention to bandpass calibration (Dwarakanath,
Carilli, \& Goss 2002).  This dynamic range is adequate to detect the
$\ge 1\%$ absorption signals discussed below.

Screening and excision of terrestrial interference is also critical to
these observations.  Much work is being done in this area in
preparation for future large area radio telescopes (Fisher 2001).  We
assume that effective techniques will be available for RFI mitigation
to the required levels.

The parameters adopted above are within the scope
of what is being considered for the SKA. 
Indeed, analyses of scientific issues such as those presented 
herein are fundamental to defining the requirements for
future large area radio telescopes. 
An interesting point to keep in mind is that
at the low frequencies considered herein,
the major cost for an SKA 
is not likely to be collecting area. Hence, 
one might consider building an even larger telescope,
thereby pushing to even fainter sources. 

\subsection{Source spectra}

The sources being considered correspond to powerful radio galaxies for
which the emission mechanism is non-thermal (synchrotron) radiation
from a relativistic plasma of electrons and magnetic fields. In most
such sources the spectrum can be described well by a
power-law over the frequency range of interest (0.6 GHz to 3 GHz in the
rest frame).  At the worst the spectra are slowly curving (in log
space) over hundreds of MHz (de Breuck et al. 2000).

In the analysis below we adopt the spectrum of the powerful radio
galaxy Cygnus A.  Cygnus A was included in the initial determination
of the absolute celestial radio flux density scale by Baars et
al. (1977), and has accurate absolute flux density measurements at
many frequencies over the frequency range of interest.  Figure 4 shows
the spectrum of Cygnus A from 500 MHz to 5000 MHz.  The solid 
line corresponds to a first order polynomial fit to the data
in the log plane, corresponding to a power-law of index
$-1.05\pm0.03$. The dashed line corresponds to a second order
polynomial. We use this second order polynomial in the analysis
below. 

\subsection{Simulated spectra for sources at $z = 8$ and 10}

Figure 5 shows a simulated spectrum at 1 kHz resolution
of a $z = 10$ radio source
with a flux density of 20 mJy at an observing frequency of
120 MHz  (S$_{120}$). The implied  luminosity density at a rest frame
frequency  of 151 MHz is then $P_{151} = 2.5\times10^{35}$ erg s$^{-1}$
Hz$^{-1}$.  Figure 5a shows a spectrum covering a large
frequency range (100 MHz to 200 MHz, or HI 21cm redshifts
of 13 to 6). Figure 5b shows an expanded view of the frequency 
range corresponding to the HI 21cm line at the source redshift
(129 MHz). 

The on-set of HI 21cm absorption by the neutral IGM is clearly
seen at 129 MHz. The general continuum level drops 
by about 1$\%$ at this frequency due to the diffuse neutral IGM.
Deeper narrow lines are also visible to frequencies as high 
as 170 MHz. Again, 
the narrow lines decrease in redshift-density with increasing
frequency. At around 130 MHz there are roughly 5 narrow lines with
$\tau \ge 0.02$ per unit MHz, while at 160 MHz the
redshift-density has decrease by a factor of 10 or so. 

Figure 6 shows a simulated spectrum at 1 kHz resolution
of a $z = 8$ radio source with  S$_{120}$ = 35 mJy, again
corresponding to $P_{151} = 2.5\times10^{35}$ erg s$^{-1}$
Hz$^{-1}$. The depression in the continuum due to absorption by
the diffuse IGM is much
less evident than at higher redshift, with a mean value of 
$\tau \sim 0.1$\%. The deep narrow lines are still easily seen,
but, again, at lower redshift-density than is found at higher
redshifts. 

\subsection{Limits to detection}

We next consider the detection limit of the absorption signal
using statistical tests. The challenge is greater at lower redshifts
due to the decreasing strength and redshift-density of the absorbing
signals, so we consider the simulated spectrum of the $z = 8$ source.  

We first consider detection of the deeper narrow lines.
The line density with redshift is such that we expect
about two to three narrow lines of  $\tau \ge 0.02$
in the range $z = 7$ to 8. In order to detect such
lines at 5$\sigma$ at the sensitivity levels considered 
herein requires a continuum source of S$_{120}$ = 11 mJy.

A second method for detecting absorption by the neutral IGM is to look
for a change in the rms noise level in the spectrum as a function of
frequency. Figure 7 shows the running rms over a spectral region of 1
MHz width at 1 kHz resolution for the simulated spectrum shown in
Figure 6 (S$_{120}$ = 35, $z = 8$).  A sharp increase of the rms noise
level above the system noise is seen starting at 157.7 MHz,
corresponding to the HI 21cm line at the source redshift. The noise
then gradually decreases back to the system value by about 185 MHz ($z
\sim 6.5$).  As a measure of the detection limit for the increased rms
due to the on-set of HI absorption by the IGM we consider a
measurement of the 'noise-on-the-noise', ie. the rms deviations,
$\sigma_{\rm rms}$, of the calculated noise values below 157.7 MHz in
Figure 7. We derive $\sigma_{\rm rms} = 0.3\mu$Jy.
These results suggest we should be able to detect the
on-set of HI 21cm absorption by the IGM at the 5$\sigma_{\rm rms}$
level toward a source with S$_{120}$ = 6.5 mJy at $z = 8$.

The third statistical test we consider is a non-parametric
Kolmogorov-Smirnov (KS)  test using cumulative distributions of the
noise. Figure 8 shows 
an example for the spectrum shown in Figure 6 ($z=8$, S$_{120}$ = 35
mJy). We calculate the noise histograms over a 10 MHz band at 1 kHz 
resolution.  The solid line is the cumulative distribution
based on the system noise, ie.  at frequencies below 157.7 MHz.
The dotted line is the distribution at frequencies just above
157.7 MHz, ie. after the on-set of HI 21cm absorption by the IGM. A
standard  KS test shows that these two distributions differ to high 
significance ($> 99\%$ quantile). Simulations for sources with lower
flux densities show that the limit to detectability (90$\%$
quantile) occurs for sources with S$_{120}$ = 12 mJy. 

These  statistical tests are certainly not 
an exhaustive analysis of the possibilities, only representative, and
the detection limits should be considered at best order-of-magnitude.

\section{Radio source populations}

The evolution of the luminosity function for
powerful radio sources is reasonable well quantified
for sources with $P_{151} \ge 6\times10^{35}$ erg s$^{-1}$ Hz$^{-1}$
out to $z \sim 4$ (Jarvis et al 2001).  
In the analysis below we must extrapolate to higher redshift,
and  lower luminosity, hence the results are necessarily speculative.

Jarvis et al (2001) derive a comoving (CM)
space density of $2.4\times 10^{-9}$ Mpc$^{-3}$ for sources with
$P_{151} \ge 6\times10^{35}$ erg s$^{-1}$ Hz$^{-1}$
at $z \sim 4$. Beyond this redshift existing surveys
are consistent with either a flat CM number density, or a steep
decline.  We consider two possibilities. The most optimistic is
the flat CM number density evolution. The second is a
steep decline in the number density with increasing $z$ 
which follows the  evolution of luminous optical QSOs as derived by
Fan et al. (2001). Fan et al. (2001) find that the CM number density
of optically luminous QSOs falls by a factor 2.5 or so per unit
redshift  beyond $z \sim 3$ to 4. While not selected as radio sources,
these optical QSOs likely correspond to accreting
massive black holes (M$_{\rm BH} \sim 10^9$; Zoltan \& Loeb 2001), 
and hence should  represent the evolution of the
massive black hole population. But even for optical QSOs
the CM number density evolution is poorly constrained 
beyond redshift 6. 

The luminosity function for powerful radio galaxies
obeys a power-law in luminosity of index $\approx 2.2$
(Jarvis et al. 2001).
The analysis in section 3.4 showed that we should be able to
detect HI 21cm absorption toward sources as faint as S$_{120} \sim 6$
mJy at $z = 8$, corresponding to $P_{151} = 4.3\times 10^{34}$ erg
s$^{-1}$ Hz$^{-1}$, or a factor 14 below the fiducial value used
by Jarvis et al. (2001).  The implied CM number density
of these sources at $z \sim 4$ is then $5.7\times 10^{-8}$ Mpc$^{-3}$.
Dunlop et al. (2002) have suggested a gross relationship 
between radio luminosity and black hole mass for powerful radio
galaxies.  While admittedly highly uncertain, their relationship 
would imply a black hole mass  of order 10$^9$ M$_\odot$ for sources
with $P_{151} = 4.3\times 10^{34}$ erg s$^{-1}$ Hz$^{-1}$.

Figure 9 shows the number  of sources in the entire sky 
with $P_{151} \ge 4.3\times 10^{34}$ erg s$^{-1}$ Hz$^{-1}$ 
per unit redshift from $z =
6$ to 15 based on the two models discussed above and normalized to the
Jarvis et al. (2001) value at $z = 4$ extrapolated to lower
luminosity.  For a flat CM number density evolution it is clear that
there will be many sources ($1.4\times 10^5$ sources
between $z = 6$ and 15)
in the sky at large enough redshift such that HI 21cm
absorption during the gray-age could be observed. In this case the
surface density of sources is a slowly decreasing function of
redshift.  Even in the case of a steeply declining source population
there will still be  a reasonable number of 
sources (2240 sources between $z = 6$ and
15) toward which absorption experiments can be performed.  In this
case the number of sources is strongly weighted to the lower
redshifts. 

\section{Discussion}

A simple argument in favor of very high-$z$ radio sources is that a
radio galaxy with a luminosity more than an order of magnitude larger
than the values discussed herein ($P_{151} = 10^{36}$ erg s$^{-1}$
Hz$^{-1}$) has already been found with a spectroscopically confirmed
redshift of $z = 5.2$ (van Breugel et al. 1999).  Extrapolating to $z
\ge 6$ seems a relatively small step in cosmic time, and below we show
that there are no obvious physical reasons to preclude radio galaxies
at these high redshifts.  The analysis in section 4 showed that
current radio surveys are at least consistent with the existence of
luminous radio sources beyond the epoch of reionization.  Indeed, 
de Breuck et al. (2002) have identified a number of candidate powerful
radio galaxies in the range $z = 6$ to 8, based on the $K - z$
relation for radio 
galaxies.  More generally, Zoltan \& Loeb (2001) show that the
formation of massive black holes ($\rm M_{BH} \le 10^9$ M$_\odot$)
starting at $z \sim 10$ is plausible in the context of hierarchical
models of structure formation.  Perhaps most compelling is the simple
fact that models of reionization require ionizing sources. Given the
already known mix of galaxies and AGN at $z \ge 6$ (Hu et al. 2002;
Fan et al. 2001), it seems likely that some of the ionizing
structures are massive accreting black holes, of which some may power
radio jets.

While the existence of massive black holes at high redshift seems
plausible, are there other physical reasons to preclude radio jets
beyond the epoch of reionization? One obvious environmental difference
is the higher density of the IGM. It is unclear if this is relevant,
at least on scales $< 100$ kpc around the black hole, since we are then
dealing with the local environment of a structure that has
already separated from the Hubble flow. 
And even if it were relevant, higher density environments
have been hypothesized to increase radio luminosities by increasing
the conversion efficiency of jet kinetic luminosity to radio
luminosity (Barthel \& Arnaud 1996).

The second obvious environmental difference at very high-$z$ is the
higher CMB temperature. Higher CMB energy densities will lead to
increased inverse Compton losses by the synchrotron emitting
relativistic electrons in the radio source.  The energy density of the
CMB increases as: $\rm U_{CMB} = 4.0\times 10^{-13} (1 + z)^4$ erg
cm$^{-3}$. The magnetic energy density in a typical extended radio
structure in high-$z$ radio galaxies is of order $\rm U_B = 8\times
10^{-10} ({{B}\over{100\mu G}})^2$, with B being the field strength in
$\mu$G.  Hence, energy losses by IC emission off the CMB are
comparable to synchrotron losses at $z \sim 6$. But radio galaxies
should be considered in the context of a 'closed box' model for
relativistic particles including active particle acceleration
(Pacholcyzk 1970). In this case one expects a steepening of the radio
spectra due to IC/synchrotron losses, not
an exponential cut-off.  In other words, the sources will have steeper
spectra, but will not be completely snuffed-out until particle
acceleration ceases. For example, the IC/synchrotron lifetime of a
relativistic electron radiating at 1.4 GHz in a 100 $\mu$G field at $z
= 7$ is of order 1 Myr. The radio emission at this, and higher,
frequencies will then be dominated by source structures where active
particle acceleration has occurred within this timescale.

Even if luminous radio sources exist during the gray-age, will
we be able to identify them?  An obvious method would
to color-select radio-loud sources in the near-IR, eg.
R-band drop-outs at $z = 7$ or  H-band dropouts at $z = 15$.
This could be done using near-IR observations of 
mJy radio samples or radio observations of near-IR samples. 
However, this method requires the objects be bright 
at (rest-frame) UV to blue wavelengths, thereby limiting the
sample to dust-poor, optically luminous sources.
A potentially more fruitful
method is to use the radio data itself. For instance, the
running-rms test could be used to detect sources with anomalously
large noise values in the relevant frequency range of 100 MHz
to 200 MHz. The initial identification could be made by comparing
the rms for a given source to the typical rms derived from all
the field sources. Once a potential high-$z$ candidate is identified,
then a second test could be done to see how the noise behaves
as a function of frequency in the candidate source spectrum. 
In this way one might
also derive the source redshift from the on-set of HI 21cm
absorption by the neutral IGM, ie. the radio Gunn-Peterson effect.

What survey area is required, and how many sources need to be
considered in order to find a radio source in the gray-age?  For the
most optimistic model (flat comoving number density evolution), the
analysis of section 4 showed that there should be about 3 sources 
deg$^{-2}$ at $z > 6.5$ with sufficient radio flux density to detect HI
21cm absorption by the IGM using the SKA.  The redshift cut-off model
based on luminous QSO evolution leads to 0.05 sources deg$^{-2}$.
The counts of celestial sources with S$_{1400} \ge 1$ mJy have been
determined by a number of groups, and all values are consistent with:
$\rm N(>S_{1400}) = (0.010\pm0.002)~ S_{1400}^{-1.0\pm0.15}$
arcmin$^{-2}$, with the 1400 MHz flux density, S$_{1400}$, in mJy
(Gruppioni et al. 1999; White et al. 1997; Windhorst et al. 1985).  At
the relevant flux density limits (S$_{1400} \sim 0.5$ mJy), the
surface density of all celestial radio sources is then about 72 sources
deg$^{-2}$.  The implied ratio of sources beyond the epoch of
reionization to foreground sources is then about $1\over{25}$ in the
flat evolution model and $1\over{1400}$ in the QSO cut-off model. In
either case it appears to be a tractable sifting problem.

It is important to emphasize that the structures giving rise to HI
21cm absorption prior to the epoch of reionization are qualitatively
different than those seen after the universe reionizes. After
reionization the HI 21cm lines arise only in rare density peaks
($\delta > 100$) corresponding to (proto)galaxies, ie. the damped Ly
$\alpha$ systems.  Prior to the epoch of fast reionization the
bulk of the IGM is neutral with a measurable opacity in the HI 21cm
line.  The absorption seen in Figure 2 arises in the ubiquitous
'cosmic web', as delineated after reionization by the Ly $\alpha$
forest (Bond, Kofman, \& Pogosian 1996).  The point is simply that the
Ly $\alpha$ forest as seen after the epoch of reionization corresponds
to structures with neutral hydrogen column densities of order
$10^{13}$ cm$^{-2}$ to $10^{15}$ cm$^{-2}$, and neutral fractions of
order $10^{-6}$ to 10$^{-4}$ (Weinberg et al. 1997).  Before
reionization these same structures will then have neutral hydrogen
column densities of order $10^{19}$ cm$^{-2}$ to $10^{20}$ cm$^{-2}$,
and hence may be detectable in HI 21cm absorption.

We conclude by emphasizing a few of the important differences between
studying the neutral IGM via HI 21cm absorption against discrete radio
sources versus via HI 21cm emission (or absorption against the
microwave background). The structures probed by absorption extend down
to galaxy scales, as opposed to the strictly large  (ie. cluster) scale
structures being probed in emission. Also, absorption features 
can be much narrower in frequency than
is expected for the emission signal (a few kHz vs. 100's of kHz,
respectively). Narrow lines are much easier to detect in the presence
of broad spectral baseline errors, as might arise from bandpass
calibration errors or from confusion due to the frequency dependent
sidelobes from the plethora of cosmic sources at low frequency.  The
sidelobe confusion problem is exacerbated by the fact that emission
studies must be done at arcminute resolution in order to have enough
material in the beam to make a detection, while absorption studies
are preferably done at high angular resolution. On the down-side,
absorption studies only probe isolated lines-of-sight, 
or at best a few lines of site in the case of an extended background
source,   and the derivation of column density is
modulo the spin temperature. The analysis in section 2 suggests that
judicious comparison of the observed absorption spectra with models of
reionization will provide significant insight into the physics of
cosmic structure formation during the gray age.  
We will address this issue in more detail in a future paper.

\vskip 0.2truein 

The National Radio Astronomy Observatory (NRAO) is operated
by Associated Universities, Inc. under a cooperative agreement with the
National Science Foundation. We thank Carlos de Breuck
for useful conversations concerning this paper.


\clearpage
\newpage

\centerline{\bf Figure Captions}

F{\scriptsize IG}. {\bf 1}.--- Evolution of the kinetic temperature of the
gas ({\it solid line\/}), the spin temperature of neutral hydrogen
({\it dashed line\/}), and the CMB temperature ({\it dotted line\/})
in the simulation.

F{\scriptsize IG}. {\bf 2}.--- The upper panel shows the transmitted
radio flux density over a relatively narrow velocity range (700 km
s$^{-1}$) assuming HI 21cm absorption by the neutral IGM.
Three different redshifts are displayed: 
$z=12$ (red), $z=10$ (violet), and $z=8$ (blue). The abscissa
for the upper panel is velocity, while that for the
middle and lower panels is the corresponding comoving physical scale. 
The middle panel shows the kinetic ({\it bold lines\/}) and the spin
({\it thin lines\/}) temperature of the 
the neutral IGM over the range of distances that contribute to the
velocity range indicated on the top panel.
The bottom panel shows
the neutral hydrogen density.  

F{\scriptsize IG}. {\bf 3}.--- The joint distribution of the spin
temperature, T$_{\rm S}$, and neutral hydrogen density, x$_{\rm HI}$(1
+ $\delta$) (in units of the mean hydrogen density) of gas along
random lines of sight at two redshifts: ({\it a\/}) $z=12$ (upper
panel) and ({\it b\/}) $z=8$ (lower panel).  The solid curves indicate
constant HI 21cm optical depth, $\tau$, as labeled.

F{\scriptsize IG}. {\bf 4}.--- The radio spectrum of the powerful
radio galaxy Cygnus A at $z = 0.057$ ($P_{151} = 1.1 \times 10^{36}$
erg s$^{-1}$ Hz$^{-1}$; Baars et al. 1977).
The dash line is a first order polynomial fit to the (log) data,
corresponding to a power-law of index $-1.05\pm 0.03$. The 
solid line is a second order polynomial fit. 

F{\scriptsize IG}. {\bf 5a}.--- The simulated spectrum from 100 MHz to
200 MHz of a source with S$_{120}$ = 20 mJy at $z = 10$ using the
Cygnus A spectral model and assuming HI 21cm absorption by the IGM.
Thermal noise has been added using the specifications of
the SKA and assuming 10 days integration with 1 kHz wide spectral
channels.  
{\bf 5b}.--- The same as 5a, but showing an expanded view of 
the spectral region around the frequency corresponding to the redshift
HI 21cm line at the source redshift (129 MHz). 
The solid line is the Cygnus A model spectrum without noise or 
absorption. 

F{\scriptsize IG}. {\bf 6a}.--- The simulated spectrum from 100 MHz to
200 MHz of a source with S$_{120}$ = 35 mJy at $z = 8$ using the
Cygnus A spectral model and assuming HI 21cm absorption by the IGM.
Thermal noise has been added using the specifications of the SKA 
and assuming 10 days integration with 1 kHz wide spectral channels. 
{\bf 6b}.--- The same as 6a, but showing an expanded view of 
the spectral region around the frequency corresponding to the redshift
HI 21cm line at the source redshift (157.8 MHz). 
The solid line is the Cygnus A model spectrum without noise or 
absorption. 

F{\scriptsize IG}. {\bf 7}.--- The rms noise of the
spectrum shown in Figure 4a ($z = 8$ source) measured in bands of 1
MHz width with 1 kHz channels. 

F{\scriptsize IG}. {\bf 8}.--- The cumulative distribution of channel
values for a 10 MHz wide band using 1 kHz channels for the spectrum in
Figure 4a ($z = 8$ source). The solid line shows the noise
distribution at frequencies just below 157.8 MHz, ie. just before the
on-set of HI 21cm absorption by the neutral IGM, while the dotted line
shows the distribution just above 157.8 MHz, ie. just after the on-set
of HI 21cm absorption by the IGM.

F{\scriptsize IG}. {\bf 9}.--- The predicted number of radio
sources with $P_{151} \ge 4.3\times 10^{34}$ erg s$^{-1}$ Hz$^{-1}$
over the entire sky per unit redshift vs. redshift using
the two models described in section 4. The normalization
is set at $z = 4$ by Jarvis et al. (2001) extrapolated to 
lower luminosity.

\clearpage
\newpage

\begin{figure}
\psfig{figure=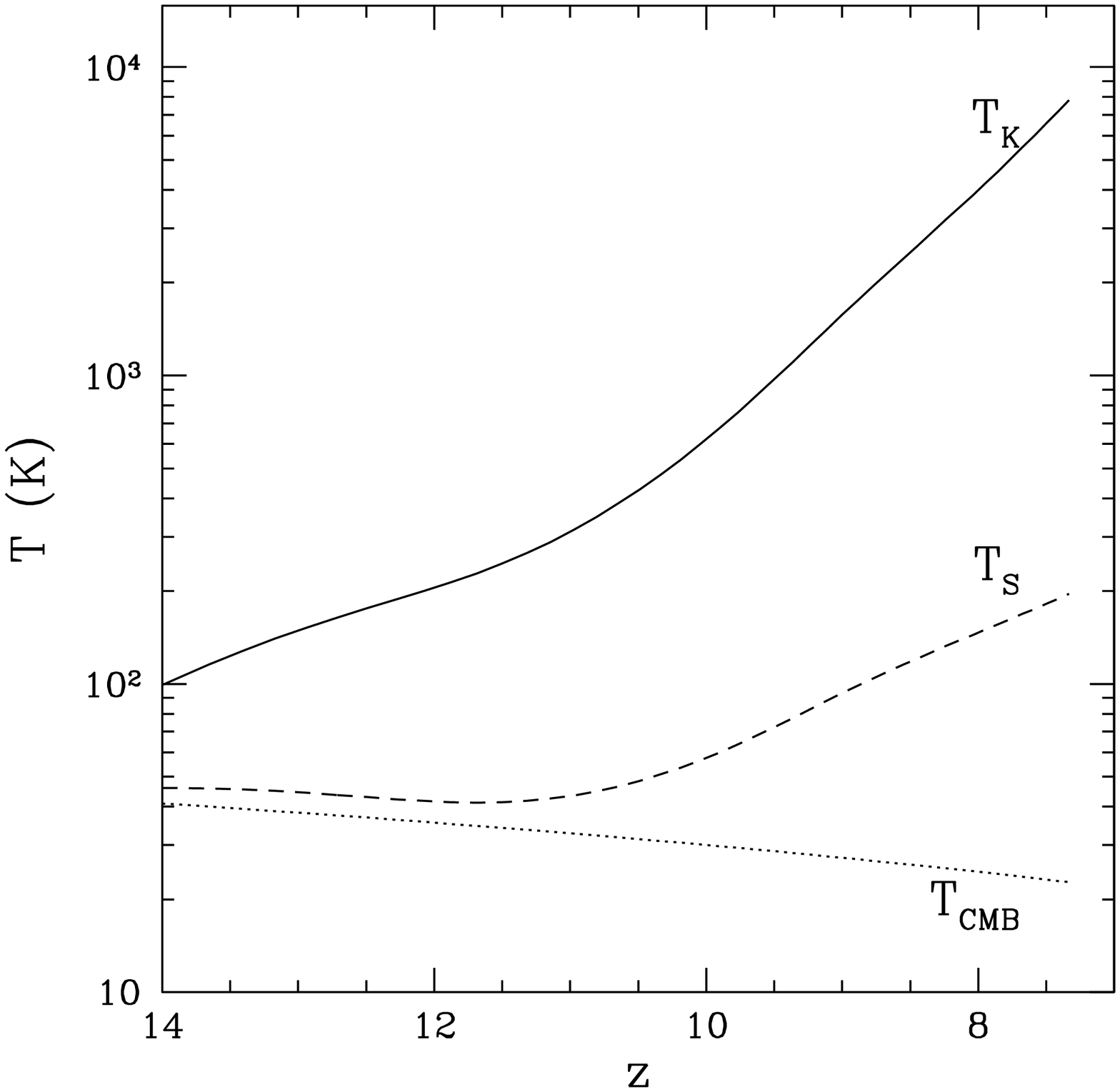,width=6in}
\caption{}
\end{figure}

\clearpage
\newpage

\begin{figure}
\psfig{figure=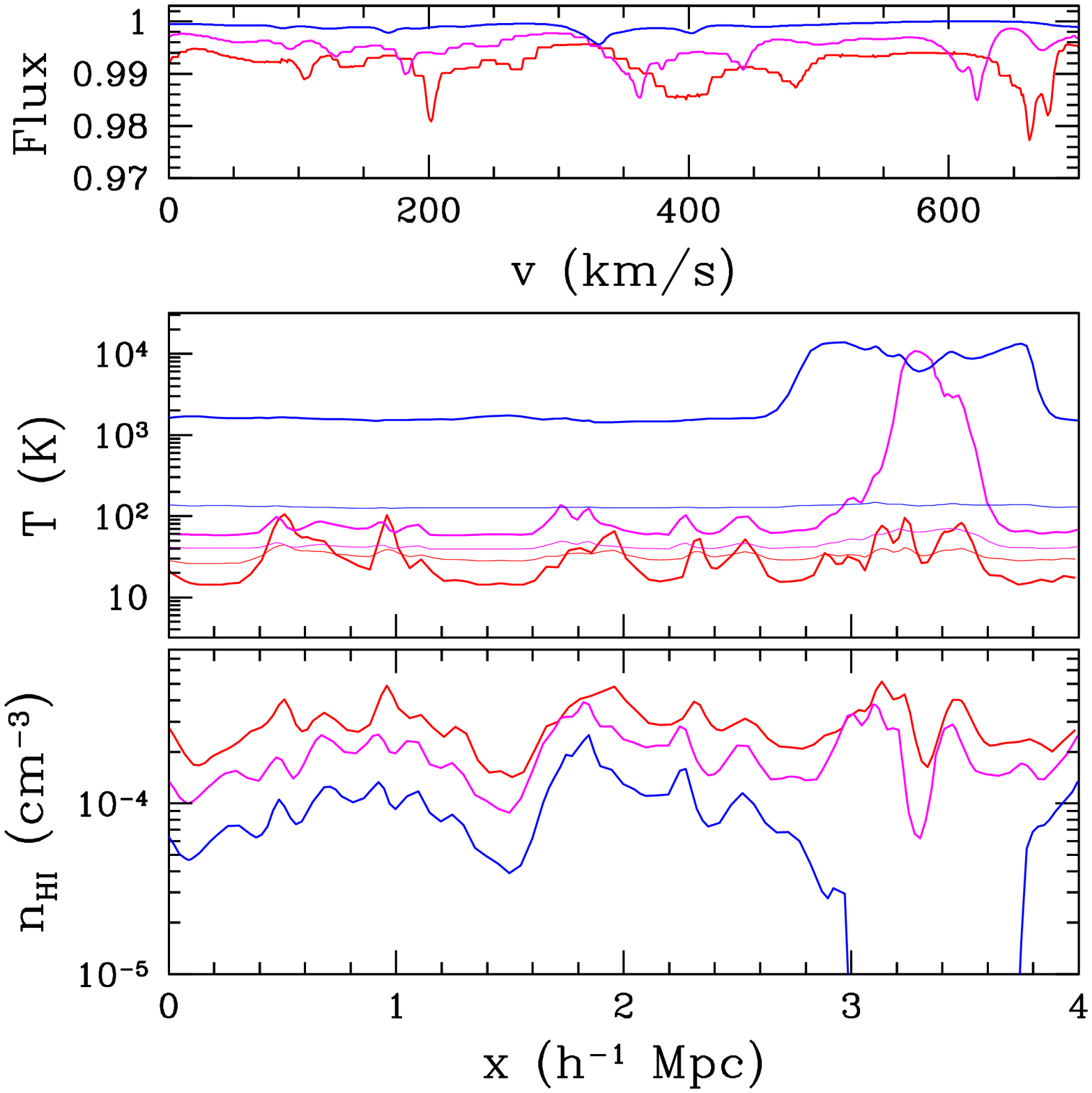,width=6in}
\caption{}
\end{figure}

\clearpage
\newpage

\begin{figure}
\vskip -1.5in
\psfig{figure=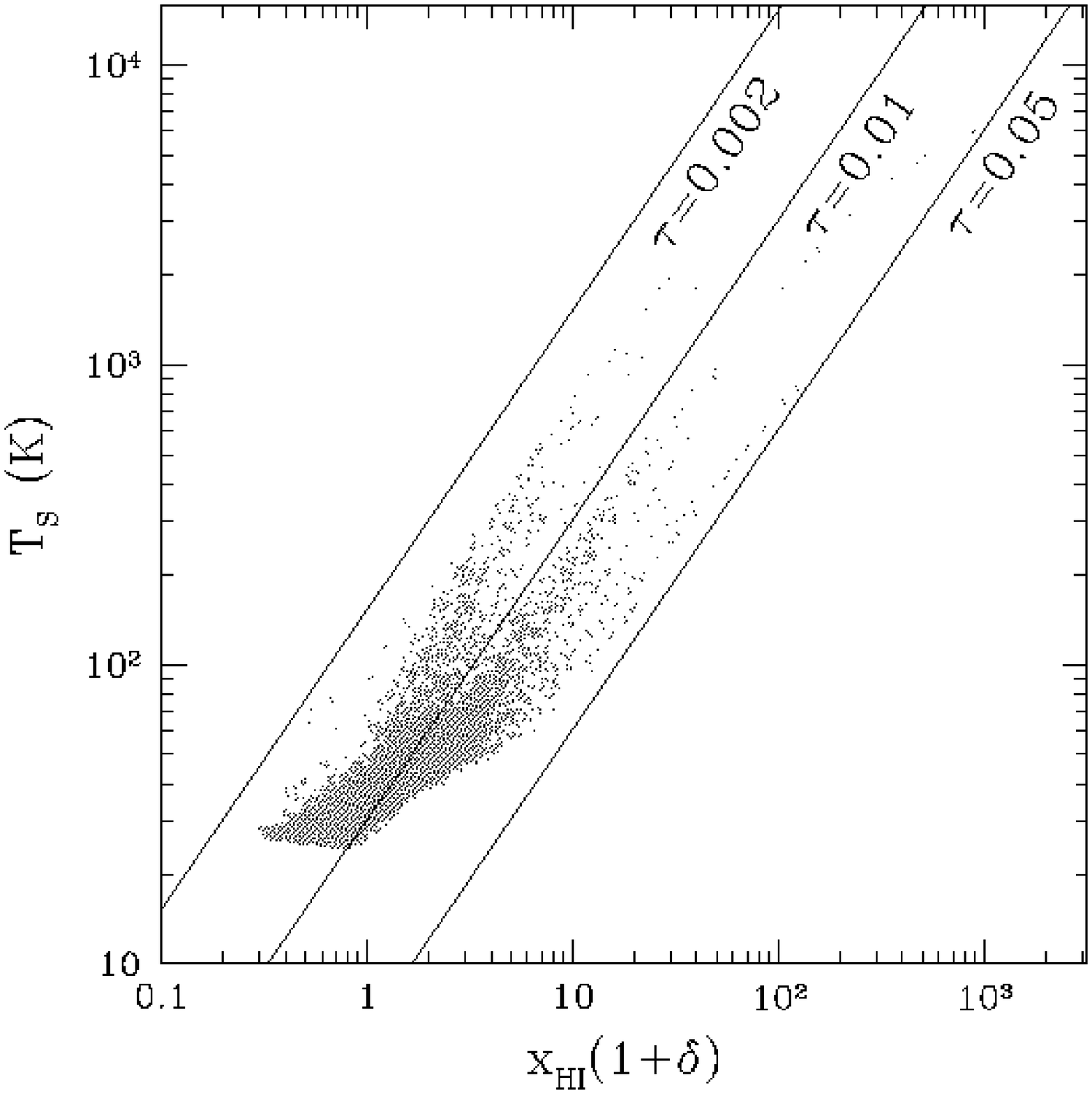,width=4in}
\psfig{figure=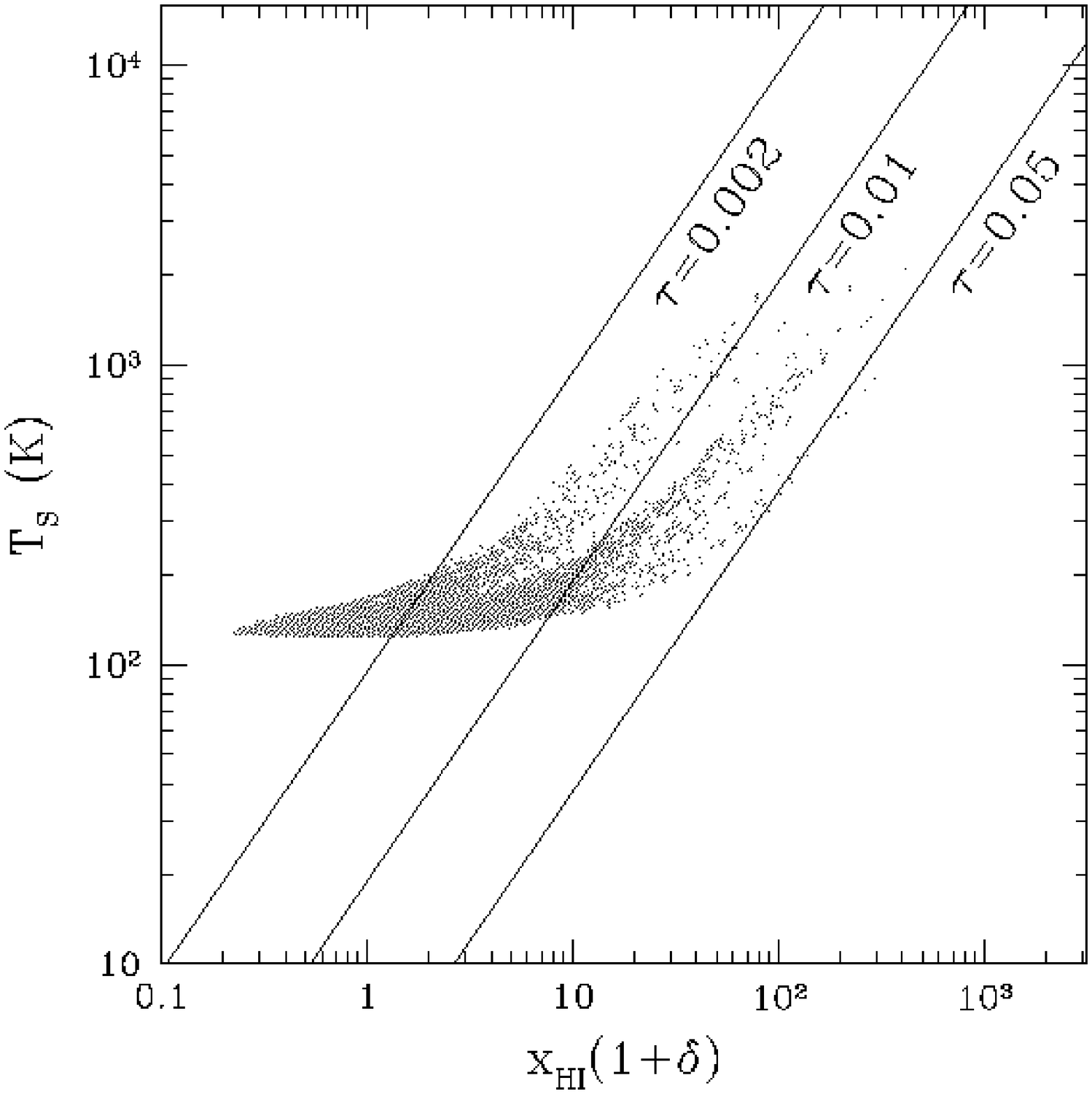,width=4in}
\caption{}
\end{figure}

\clearpage
\newpage

\begin{figure}
\psfig{figure=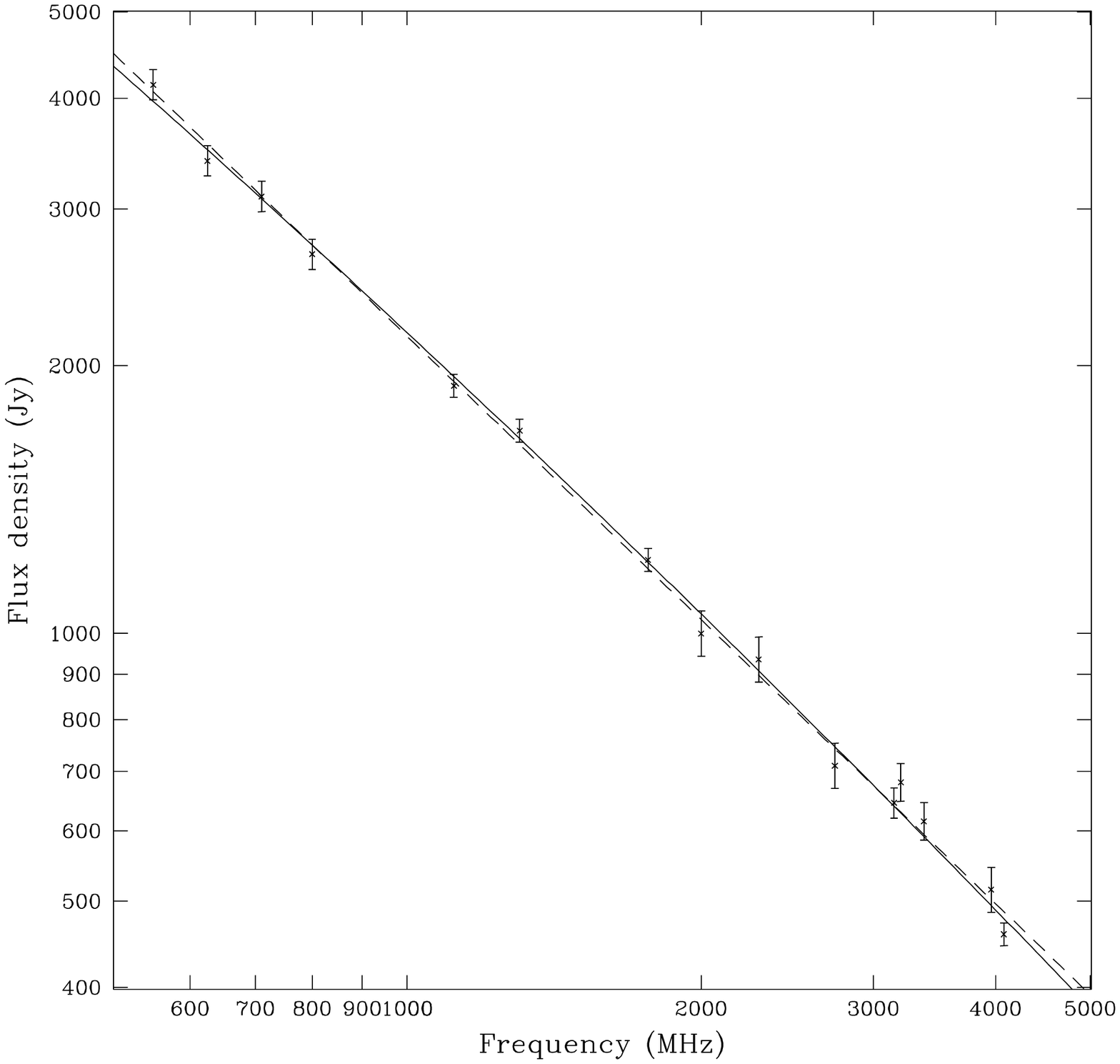,width=6in}
\caption{}
\end{figure}

\clearpage
\newpage

\begin{figure}
\vskip -1.5in
\psfig{figure=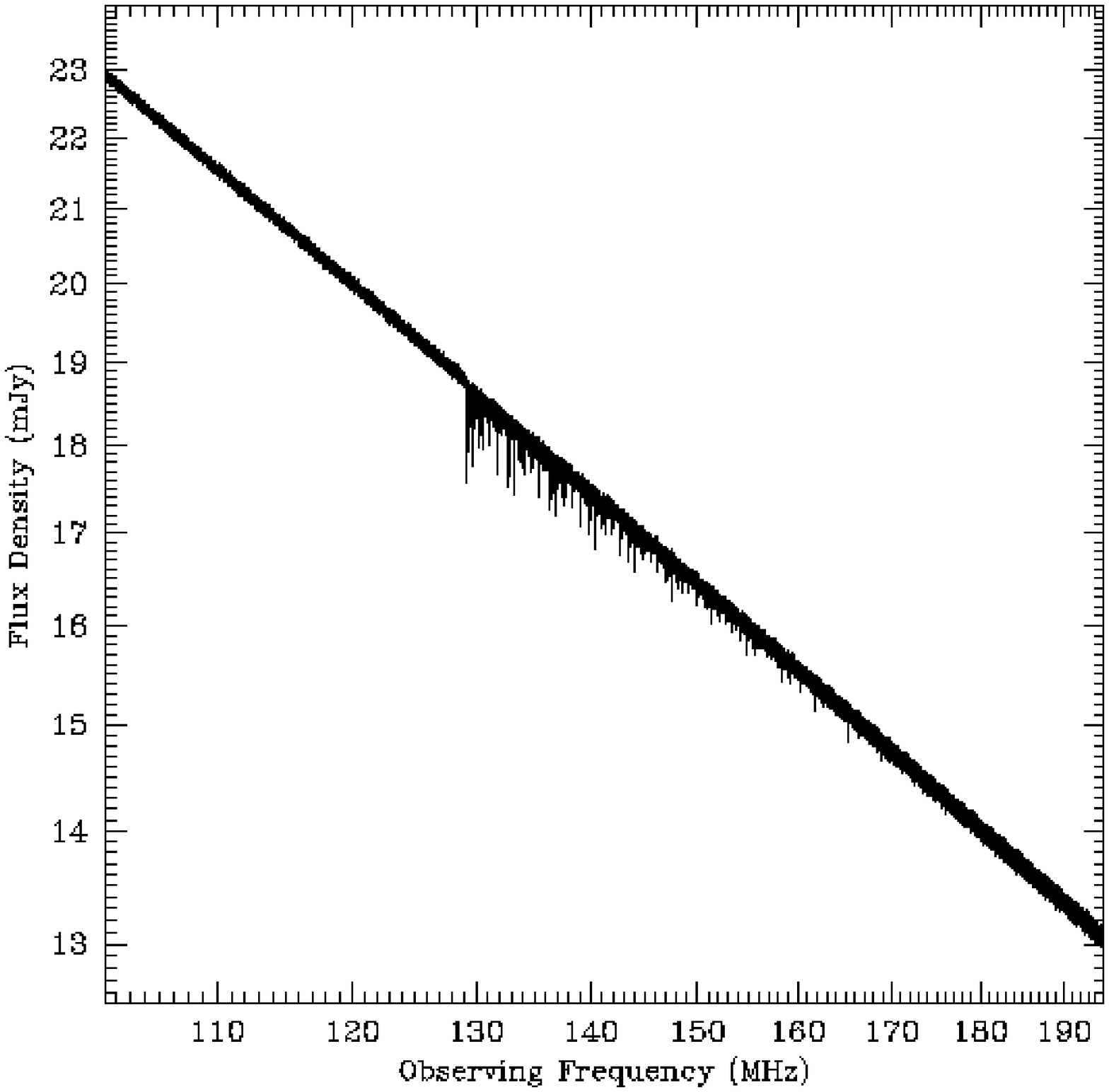,width=4in}
\psfig{figure=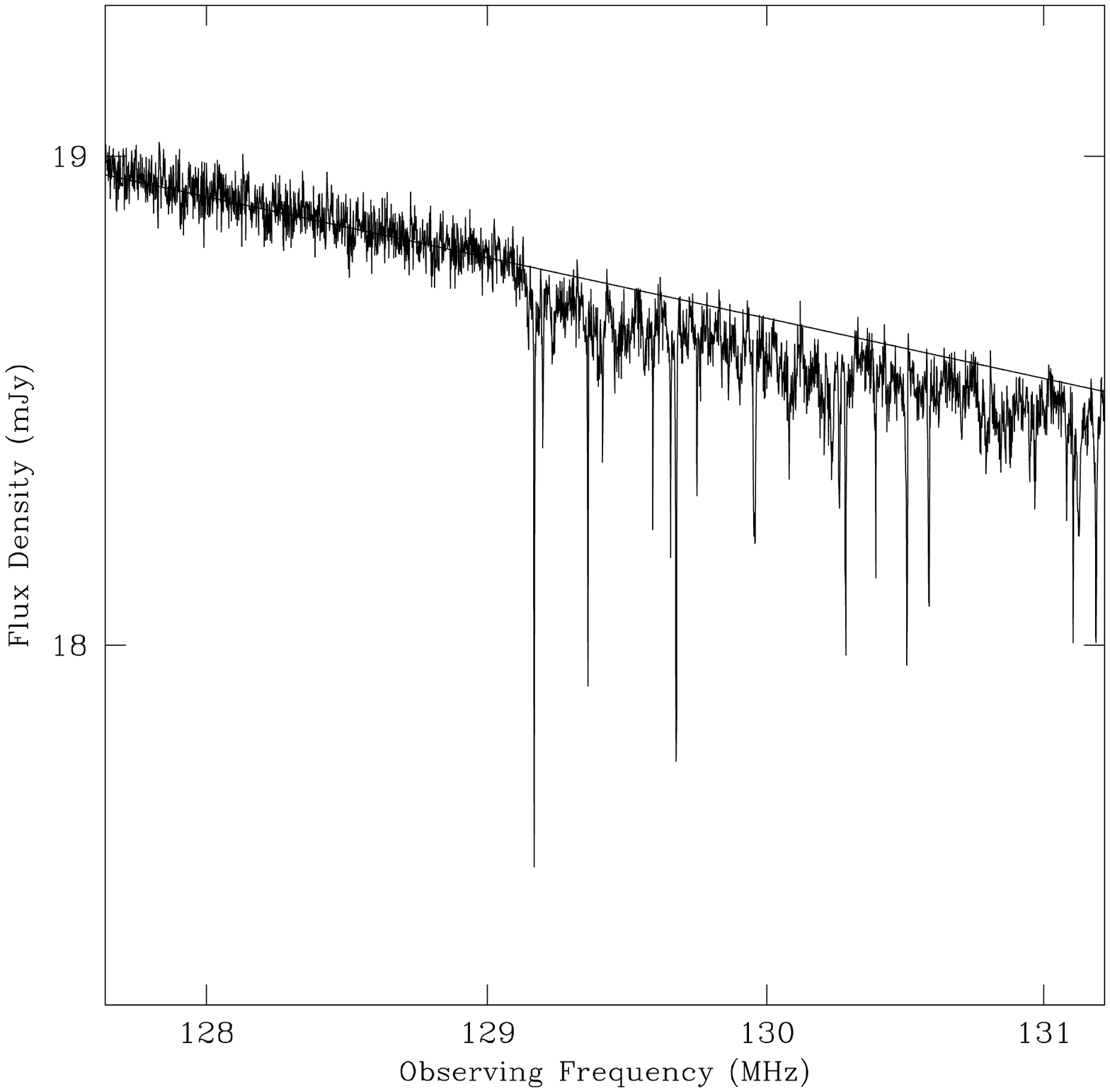,width=4in}
\caption{}
\end{figure}

\clearpage
\newpage

\begin{figure}
\vskip -1.5in
\psfig{figure=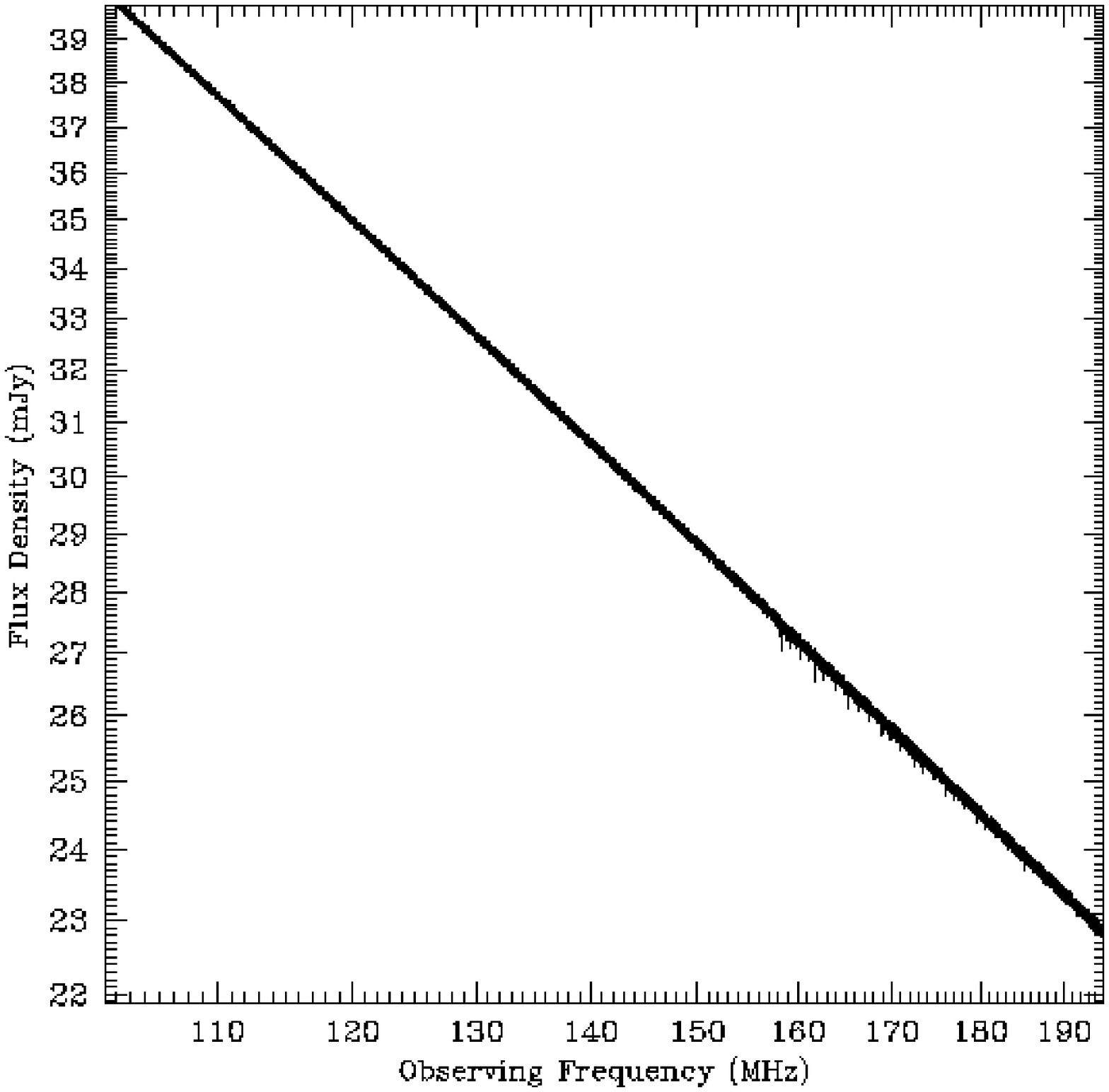,width=4in}
\psfig{figure=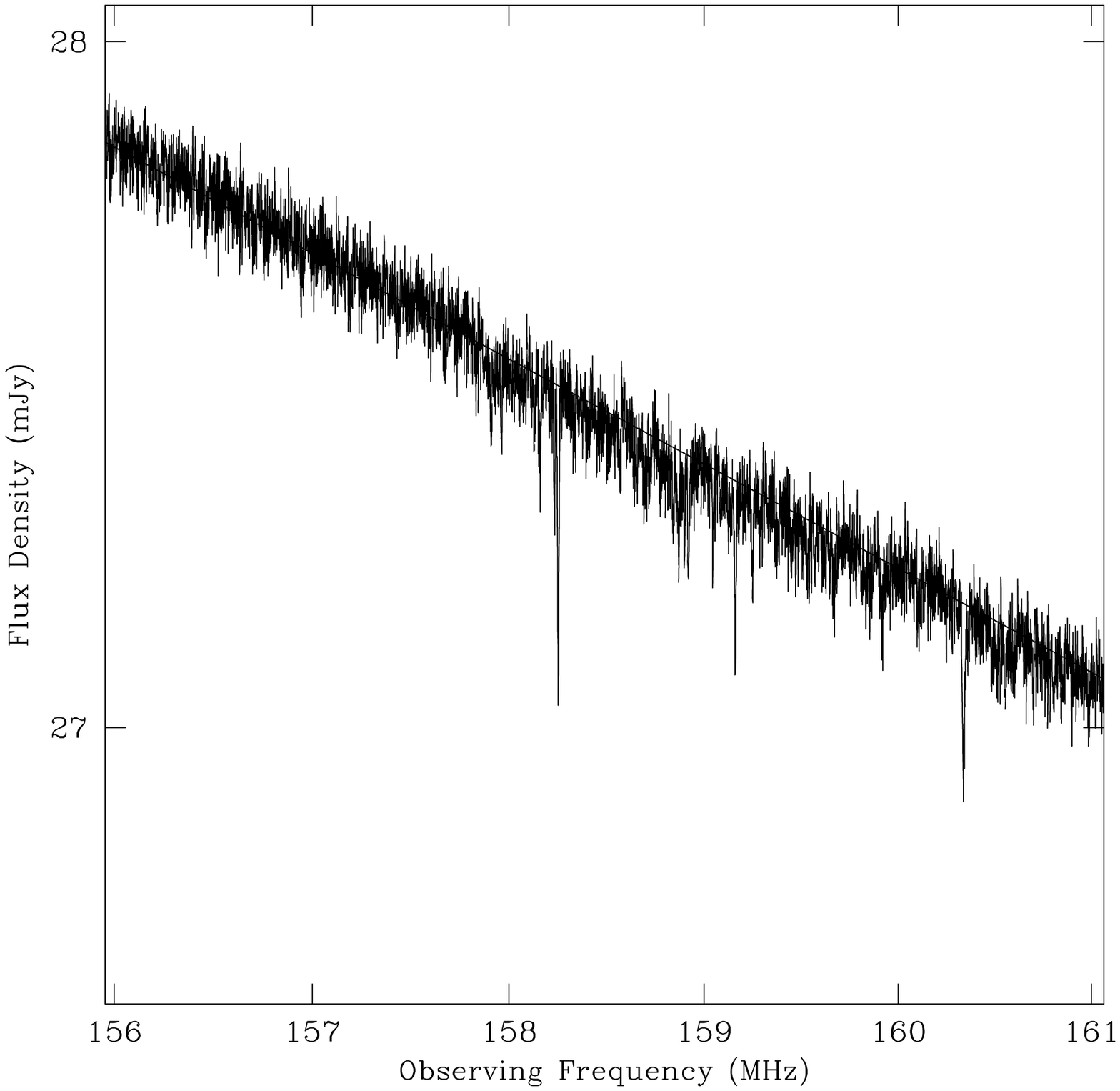,width=4in}
\caption{}
\end{figure}

\clearpage
\newpage

\begin{figure}
\psfig{figure=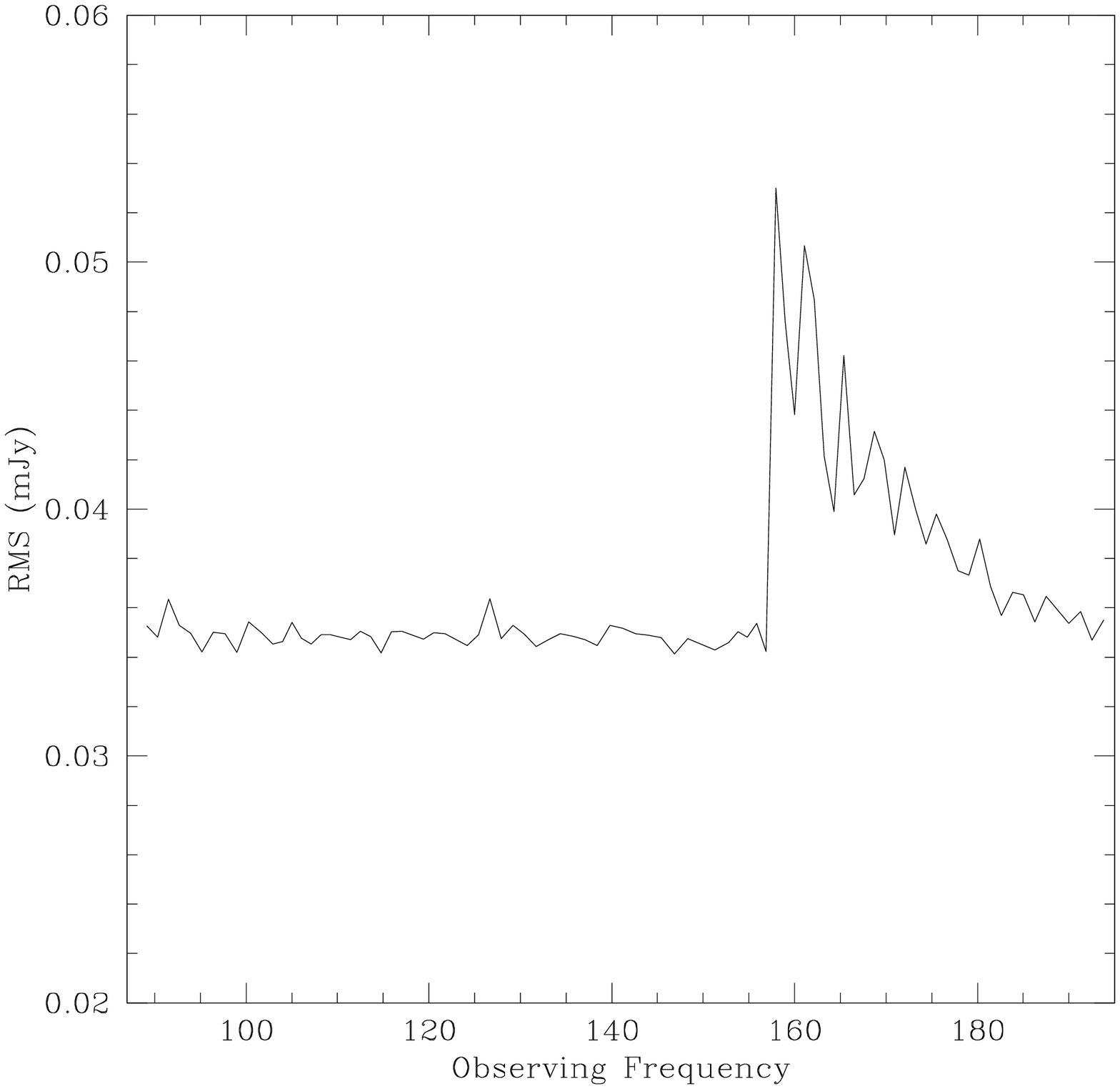,width=6in}
\caption{}
\end{figure}

\clearpage
\newpage

\begin{figure}
\psfig{figure=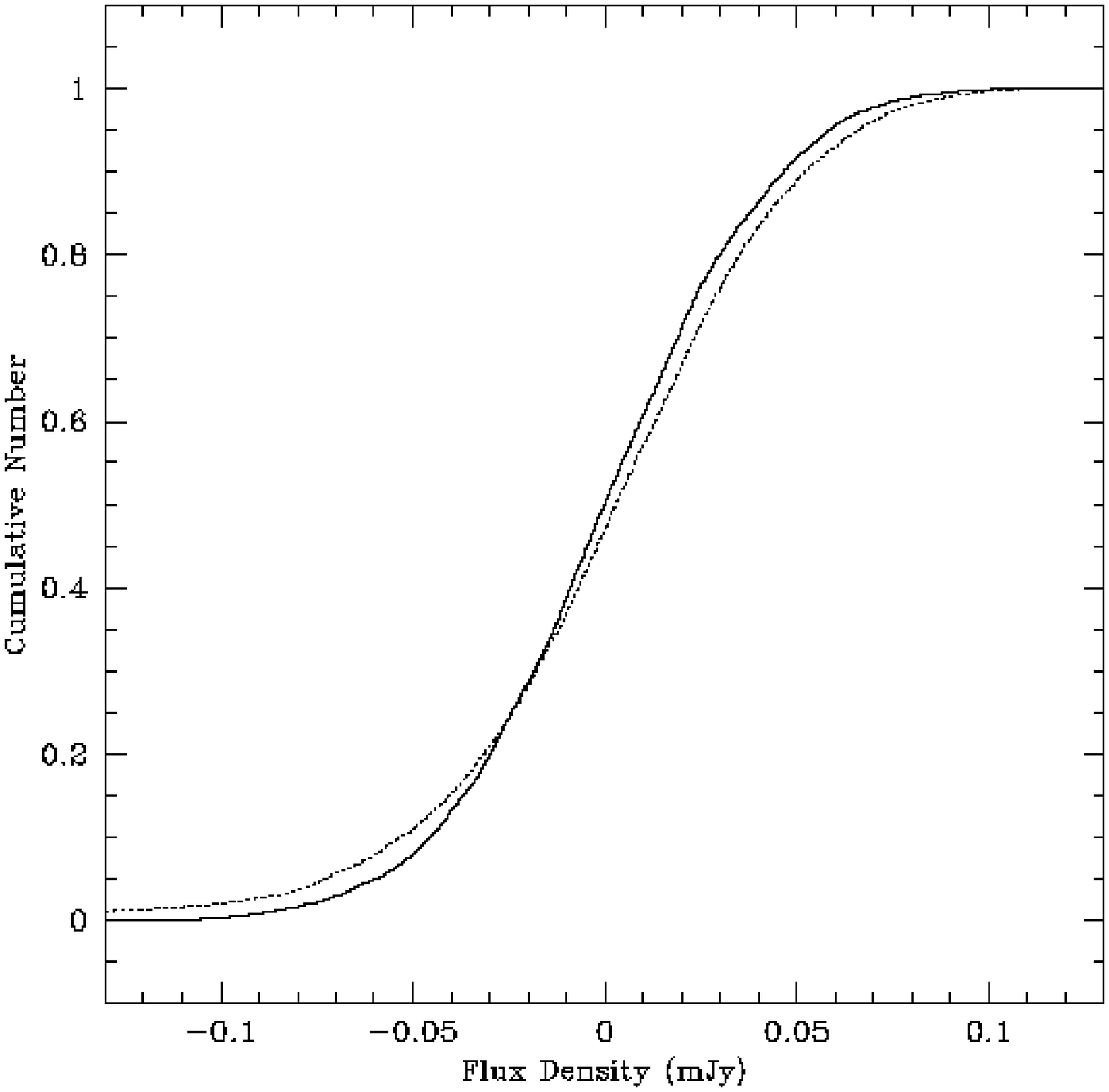,width=6in}
\caption{}
\end{figure}

\clearpage
\newpage

\begin{figure}
\psfig{figure=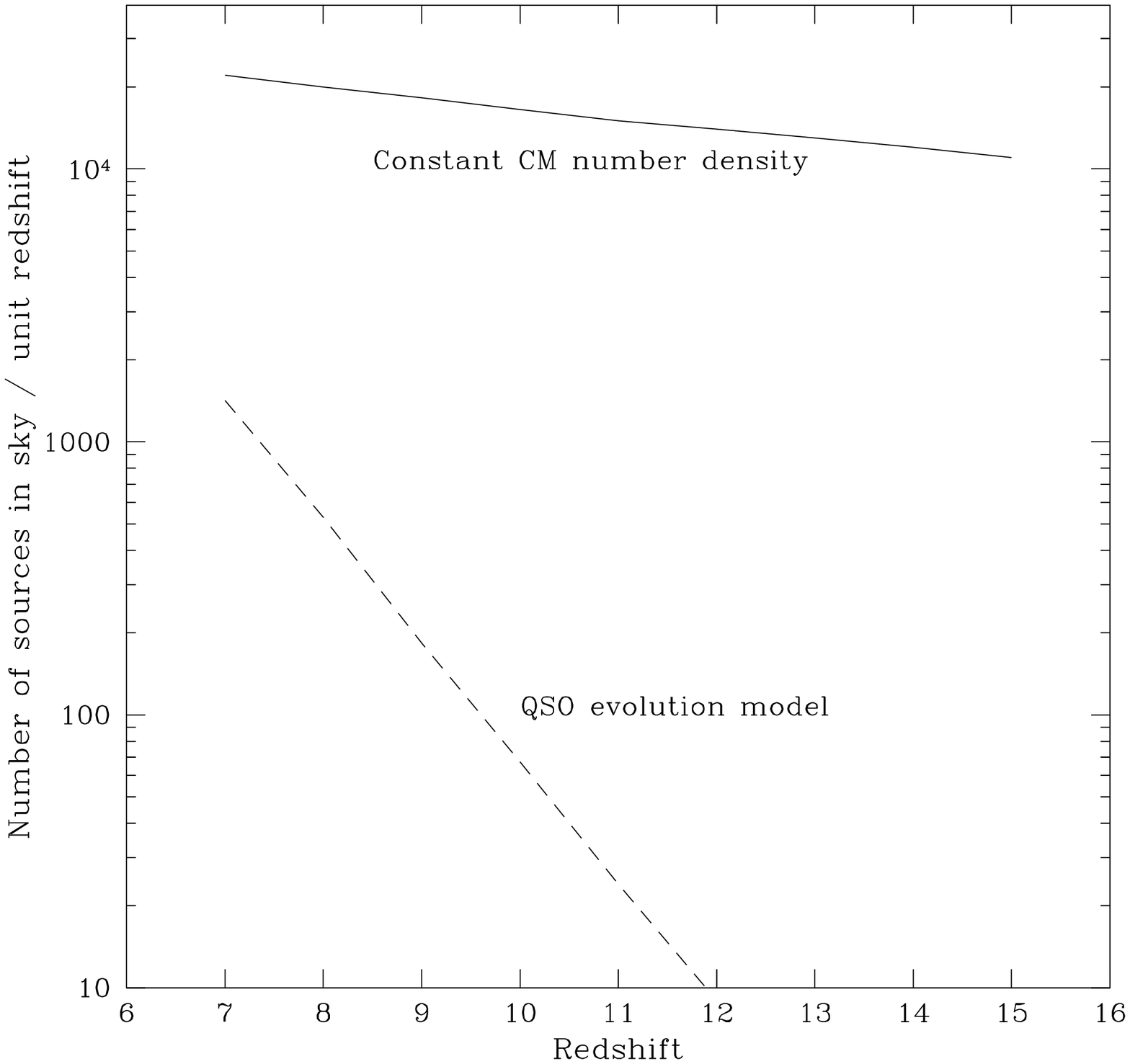,width=6in}
\caption{}
\end{figure}

\clearpage
\newpage

\end{document}